# Development and validation of a sequence of clicker questions for helping students learn addition of angular momentum in quantum mechanics

Paul Justice, Emily Marshman, and Chandralekha Singh

*Department of Physics and Astronomy, University of Pittsburgh, Pittsburgh, PA, 15260*

**Abstract.** Engaging students with well-designed clicker questions is one of the commonly used research-based instructional strategy in physics courses partly because it has a relatively low barrier to implementation [1]. Moreover, validated robust sequences of clicker questions are likely to provide better scaffolding support and guidance to help students build a good knowledge structure of physics than an individual clicker question on a particular topic. Here we discuss the development, validation and in-class implementation of a clicker question sequence (CQS) for helping advanced undergraduate students learn about addition of angular momentum, which takes advantage of the learning goals and inquiry-based guided learning sequences in a previously validated Quantum Interactive Learning Tutorial (QuILT). The in-class evaluation of the CQS using peer instruction is discussed by comparing upper-level undergraduate students' performance after engaging with the CQS with previous published data from the QuILT pertaining to these concepts.

## I. INTRODUCTION AND BACKGROUND

Clicker questions (also known as concept tests) are conceptual multiple-choice questions typically administered in the classroom to engage students in the learning process and obtain feedback about their learning via a live feedback system called clickers [1-2]. Integration of peer interaction with lectures via clicker questions has been popularized in the physics community by Mazur [2]. In Mazur's approach, the instructor poses conceptual, multiple-choice clicker questions to students which are integrated throughout the lecture. Students first answer each clicker question individually, which requires them to take a stance regarding their thoughts about the concept(s) involved. Students then discuss their answers to the questions with their peers and learn by articulating their thought processes and assimilating their thoughts with those of the peers. Then after the peer discussion, they answer the question again using clickers followed by a general class discussion about those concepts in which both students and the instructor participate. The feedback that the instructor obtains is also valuable because the instructor has an estimate of the prevalence of student common difficulties and the fraction of the class that has understood the concepts and can apply them in the context in which the clicker questions are posed. The use of clickers keeps students alert during lectures and helps them monitor their learning. Clicker questions can be used in the classroom in different situations, e.g., they can be interspersed within lectures to evaluate student learning in each segment of a class focusing on a concept, at the end of a class or to review materials from previous classes at the beginning of a class.

While clicker questions for introductory [2] and upper-level physics such as quantum mechanics [3] have been developed, there have been very few efforts [4] toward a systematic development and validation of clicker question sequences (CQSs), e.g., question sequences on a given concept that can be used in a few class periods when students learn the concepts and that build on each other effectively to help students organize their knowledge structure. Here we discuss the development, validation and implementation of a CQS on addition of angular momentum in quantum mechanics (QM) that was developed for students in upper-level undergraduate QM courses by taking advantage of the learning goals and inquiry-based guided learning sequences in a research-validated Quantum Interactive Learning Tutorial (QuILT) on this topic [5] as well as by refining, fine-tuning and adding to the existing clicker questions from our group which have already been individually validated [3]. The CQS can be used in class either separately from the QuILT or synergistically with the corresponding QuILT [5] if students engage with the QuILT after the CQS as another opportunity to reinforce the concepts learned.

## II. LEARNING GOALS AND METHODOLOGY

The learning goals and inquiry-based learning sequences in the QuILT, which guided the development and sequencing of the CQS questions, were developed using extensive research on student difficulties with these concepts as a guide and cognitive task analysis from expert perspective.

**Learning Goals**: One learning goal of the CQS (consistent with the QuILT) is that students should be able to identify the dimensionality of the product space of the spin of two particles. For example, if a system consists of two spin-1 particles with individual three-dimensional spin Hilbert spaces, the product space of the two spin system is the product of those dimensions, 3×3=9 (not the sum of dimensions, 3+3=6). Another learning goal of the CQS is that students are able to choose a suitable representation, such as the "uncoupled" or "coupled" representation, and construct a complete set of basis states for the product space in that representation. We note that the concepts related to the addition of orbital and spin angular momenta are analogous so here we will only focus on spin. In standard notation, the basis states in the uncoupled representation are eigenstates of $\hat{S}_1^2$, $\hat{S}_{z1}$, $\hat{S}_2^2$ and $\hat{S}_{z2}$ and can be written as $|s_1, m_{s1}\rangle \otimes |s_2, m_{s2}\rangle$. Here each particle's individual spin or z-component of spin quantum numbers are $s_1$, $s_2$ and $m_{s1}$,

$m_{s2}$, respectively. On the other hand, in the coupled representation, the basis states, $|s, m_s\rangle$, are eigenstates of $\hat{S}^2$ and $\hat{S}_z$ where $\hat{\vec{S}} = \hat{\vec{S}}_1 + \hat{\vec{S}}_2$ and the total spin quantum number, s, and the z-component of the spin quantum number, $m_s$, are for the entire system. Students should be able to use the addition of angular momentum to determine that the total spin quantum number of the system s can range from $s_1 + s_2$ down to $|s_1 - s_2|$, with integer steps in between, where $s_1$ and $s_2$ are the individual spin quantum numbers for the particles. The z-component of the spin of the composite system is $m_s = m_{s1} + m_{s2}$. Another learning goal of the CQS is that students are able to calculate matrix elements of various operators corresponding to observables (e.g., a Hamiltonian in product space) in different representations.

**Development and Validation**: Based upon the learning goals delineated in the QuILT, questions in the addition of angular momentum CQS were developed or adapted from prior validated clicker questions and sequenced to balance difficulties, avoid change of both concept and context between adjacent questions as appropriate in order to avoid a cognitive overload, and include a mix of abstract and concrete questions to help students develop a good grasp of the concepts. The validation was an iterative process.

After the initial development of the additional of angular momentum CQS using the learning goals and inquiry-based guided sequences in the QuILT and existing individually validated CQSs, we iterated the CQS with three physics faculty who provided valuable feedback on fine-tuning and refining both the CQS as a whole and some new questions that were developed and adapted with existing ones to build the CQS to ensure that the questions were unambiguously worded and build on each other based upon the learning goals. We then conducted individual think-aloud interviews with advanced students who had learned these concepts via traditional lecture-based instruction in relevant concepts to ensure that they interpreted the CQS questions as intended and the sequencing of the questions provided the appropriate scaffolding support to students. The final version of the CQS has 11 questions, which can be grouped into three sections (to be discussed below), and can be integrated with lectures in which these relevant concepts are covered in a variety of ways based upon the instructor's preferences.

The addition of angular momentum CQS has three sections that can be used separately or together depending, e.g., upon whether these are integrated with lectures similar to Mazur's approach, used at the end of each class or used to review concepts after students have learned via lectures everything related to addition of angular momentum that the instructor wanted to teach. The first section of the CQS, CQ1-CQ3, focuses on the uncoupled representation with basis states $|s_1 m_{s1}\rangle \otimes |s_2, m_{s2}\rangle$. The first question focuses on student understanding of the notation for the basis states in this representation along with the dimensionality of the product space and what a complete set of basis states looks like. Following this question, CQ2 and CQ3 build on this understanding, asking students to identify the operators for which the basis states in the uncoupled representation are eigenstates and about some diagonal and off-diagonal matrix elements of various operators and whether they are zero or non-zero (i.e., determining whether operators are diagonal in the uncoupled representation). This section of the CQS concludes with a class discussion in which the instructor may review characteristics of this representation, as well as address any common difficulties exhibited by students.

The second section of this CQS, CQ4-CQ6, deals with the coupled representation with basis states $|s, m_s\rangle$. The structure and concepts in these questions shown below are analogous to the structure of the first section, allowing students to compare and contrast these two representations.

*(CQ4)* Choose all of the following statements about the product space for a system of **two spin-1/2 particles** in the **coupled representation** that are correct:
I. The dimensionality of the product space is the product of the dimensions of each particle's subspace, which is 2x2=4.
II. $|s, m_s\rangle$ is an appropriate form for the basis states, where s ranges from $|s_1-s_2|$ to $s_1+s_2$ by integer steps, and $m_s=m_{s1}+m_{s2}$, ranging from –s to s in integer steps for each s.
III. $|1,1\rangle$, $|1,0\rangle$, $|1,-1\rangle$, and $|0,0\rangle$ are the elements of a complete set of basis states.
  a) I only    b) I and II only
  c) I and III only    d) II and III only
  **e) All of the above**

*(CQ5)* Choose all of the following statements about the product space for a system of **two spin-1/2 particles** in the **coupled representation** that are correct:
I. Basis state $|1,-1\rangle$ is an eigenstate of $\hat{S}^2$ such that $\hat{S}^2|1,-1\rangle = 2\hbar^2|1,-1\rangle$.
II. Basis state $|1,-1\rangle$ is an eigenstate of both $\hat{S}_1^2$ and $\hat{S}_2^2$ such that $\hat{S}_1^2|1,-1\rangle = 2\hbar^2|1,-1\rangle$ and $\hat{S}_2^2|1,-1\rangle = 2\hbar^2|1,-1\rangle$.
III. Basis state $|1,-1\rangle$ is an eigenstate of $\hat{S}_{z1}$, $\hat{S}_{z2}$, and $\hat{S}_z$.
  **a) I only**    b) I and II only
  c) I and III only    d) II and III only
  e) All of the above

*(CQ6)* Consider the product space of a system of **two spin-1/2 particles**. Choose all of the following that are correct regarding the scalar products in the coupled representation. (Recall that these scalar products give the matrix elements of the $\hat{S}_{1z} + \hat{S}_{2z}$ operator in this basis).
I. $\langle 1,1|(\hat{S}_{z1} + \hat{S}_{z2})|1,0\rangle = \langle 1,1|\hat{S}_z|1,0\rangle = 0$
II. $\langle 1,-1|(\hat{S}_{z1} + \hat{S}_{z2})|1,-1\rangle = \langle 1,-1|\hat{S}_z|1,-1\rangle = -\hbar$
III. $(\hat{S}_{z1} + \hat{S}_{z2})$ is diagonal in the coupled representation.
IV. $(\hat{S}_{z1} + \hat{S}_{z2})$ is diagonal in the uncoupled representation.
  a) II and III only    b) I, II, and III only
  c) I and IV only    d) I, II, and IV only
  **e) All of the above**

As noted, the first two sections of the addition of angular momentum CQS deal with only one representation at a time, and only with a system of two spin-1/2 particles. This choice is deliberate by design to avoid cognitive overload and allow students to revisit these representations in familiar context since typical instruction on these concepts tends to emphasize a system of two spin-1/2 particles first.

The third section of the CQS extends these concepts to higher dimensional product spaces for both coupled and uncoupled representations. For example, CQ7 deals with the dimensionalities of the product space for systems of two spins that are not both spin-1/2. Then, CQ8 and CQ9 ask students to identify basis states in the coupled and uncoupled representations for these less familiar two-spin systems. Finally, CQ10 and CQ11 ask students to identify the basis in the product space in which given Hamiltonians are diagonal. These Hamiltonians are comprised of operators addressed previously in the first questions of the CQS.

**In-Class Implementation**: The CQS was implemented with peer discussion [2-3] in an upper-level undergraduate QM class at a large research university (Pitt) after traditional lecture-based instruction in relevant concepts on the addition of angular momentum in which students learned about the coupled and uncoupled representations not only for a system of two spin-1/2 particles but also for systems for which the product spaces involve higher dimensions. Prior to the implementation of the CQS in class, students took a pretest after traditional instruction, which was developed and validated by Zhu et al. [5] to measure comprehension of the concepts of addition of angular momentum. The first six questions in the CQS were implemented together right after the pretest and the last five questions in the third section of the addition of angular momentum CQS were implemented at the beginning of the next class to review concepts covered earlier in the lectures on product spaces involving higher dimensions. The posttest was administered during the following week to measure the impact of the CQS.

In the pretest, in standard notations, students were given a system of two spin-1/2 particles and a spin-spin interaction Hamiltonian, $\widehat{H}_1 = \left(4E_0/\hbar^2\right)\hat{S}_1 \cdot \hat{S}_2 = \left(2E_0/\hbar^2\right)(\hat{S}^2 - \hat{S}_1^2 - \hat{S}_2^2)$, and a magnetic field-spin interaction Hamiltonian, $\widehat{H}_2 = -\mu B(\hat{S}_{z1} + \hat{S}_{z2})$ and asked to answer these questions:

**(a)** *Write down a complete set of basis states for the product space of a system of two spin-1/2 particles. Explain the labels you are using to identify your basis states.*

**(b)** *Evaluate one diagonal and one off-diagonal matrix element of the Hamiltonian $\widehat{H}_1$ (of your choosing) in the basis you have chosen. Label the matrix elements so that it is clear which matrix elements they are.*

**(c)** *Evaluate one diagonal and one off-diagonal matrix element of the Hamiltonian $\widehat{H}_2$ (of your choosing) in the basis you have chosen. Label the matrix elements so that it is clear which matrix elements they are.*

**(d)** *Are both Hamiltonians $\widehat{H}_1$ and $\widehat{H}_2$ diagonal matrices in the basis you chose?*

The posttest that students were administered following the implementation of the CQS was analogous to the pretest [5] and asked the same questions as the pretest but for a system of one spin-1/2 particle and one spin-1 particle. These pre-/posttests are very similar to those administered by Zhu et al. to measure student learning after traditional instruction and after engaging with the addition of angular momentum QuILT [5]. However, due to time constraints in the classroom, questions (b) and (c), which had previously asked students to construct the entire matrix representation of the Hamiltonians, were reduced as stated earlier to evaluation of only one diagonal and off-diagonal matrix element [5]. In order to compare the performance of CQS and QuILT groups on pre-/posttests so that the relative improvements can be determined, the same rubric was used for pre-/posttests given to the CQS students as the QuILT students in Ref. [5] (who were also advanced undergraduate students in QM). Questions (a), (b), and (c) were each worth 3 points, and students were awarded partial credit if only some basis states in (a) or some matrix elements in (b) or (c) were correct. Question (d) was worth 1 point (correct answer "yes or no").

## III. IN-CLASS IMPLEMENTATION RESULTS

Tables 1 and 2 compare pre-/posttest performances of upper-level QM students from the same university in two different years after traditional lecture-based instruction (pretest) and on posttest after students had engaged with the CQS (Table 1) or QuILT (Table 2) on the addition of angular momentum. The normalized gain (or gain) is calculated as $g = (post\% - pre\%)/(100\% - pre\%)$ [6] and presented in both Tables 1 and 2 but effect size is calculated only in Table 2 (not available for Table 1 data in Ref. [5]). Effect size was calculated as Cohen's $d = (\mu_{post} - \mu_{pre})/\sigma_{pooled}$ where $\mu_i$ is the mean of group $i$ and where the pooled standard deviation is $\sigma_{pooled} = \sqrt{\sigma_{pre}^2 + \sigma_{post}^2}$ [7].

**Table 1.** Comparison of mean pre/posttest scores on each question, normalized gains and effect sizes for upper-level undergraduate QM students who engaged with the CQS on addition of angular momentum concepts (N=16).

| Part | Pretest Mean | Posttest Mean | Normalized Gain (g) | Effect Size (d) |
|---|---|---|---|---|
| (a) | 59% | 95% | 0.88 | 0.30 |
| (b) | 24% | 48% | 0.31 | 0.22 |
| (c) | 17% | 71% | 0.66 | 0.44 |
| (d) | 14% | 43% | 0.33 | 0.67 |

**Table 2.** Comparison of mean pre/posttest scores on each question and normalized gains from Ref. [5] (effect sizes not available) for upper-level undergraduate QM students who engaged with the QuILT on addition of angular momentum concepts (N=26).

| Part | Pretest Mean | Posttest Mean | Normalized Gain (g) |
|---|---|---|---|
| (a) | 77% | 85% | 0.35 |
| (b) | 8% | 54% | 0.50 |
| (c) | 8% | 73% | 0.71 |
| (d) | 31% | 85% | 0.78 |

Although the number of students in each class is small and the pretest scores in Tables I and II are often different, they are low in both tables (except for question (a) in Table 2). However, the comparison of the posttest scores of the CQS group and the QuILT group in Tables I and II suggests that the CQS is effective in helping students learn to construct a complete set of basis states (question (a)) and calculate matrix elements for the magnetic field-spin interaction Hamiltonian (question (c)), garnering similar posttest scores to those of students who engaged with the QuILT. However, Table 1 also shows that students did not perform well on questions (b) and (d) even after engaging with the CQS. Analysis of data suggests that a major reason for the poor performance on both of these questions even after the CQS is due to the fact that a majority of students chose the basis to be uncoupled representation (since it is the simpler representation for constructing the basis states) and then had difficulty with the matrix elements of the spin-spin interaction Hamiltonian in question (b) since it is only diagonal in the coupled representation. In particular, in question (a), many students correctly constructed a complete set of basis states, but chose the uncoupled representation.

We note that while the magnetic field-spin interaction Hamiltonian in question (c) is diagonal in both coupled and uncoupled representations, calculating the matrix elements of the spin-spin interaction Hamiltonian in question (b) in the uncoupled representation is challenging since that operator is not diagonal in this basis. Along with the good posttest score for question (a), the CQS group students' poor posttest score on questions (b) and (d) in Table 1 is due to the fact that while students learned to construct a complete set of basis states, many were not versed in calculating the matrix elements of an operator in a representation in which it is not diagonal as in question (b) (many students assumed that the spin-spin Hamiltonian in question (b) is also diagonal in the uncoupled representation, which it is not).

In fact, for question (d), even after the CQS, many students claimed that both Hamiltonians are diagonal in the uncoupled representation they had chosen. Since students were only asked to calculate a single off-diagonal matrix element in question (b), some students who correctly calculated an off-diagonal matrix element in question (b) that was zero concluded that the entire $\hat{H}_1$ matrix is diagonal in the uncoupled representation which it is not. On the other hand, a comparison of student performances on posttest in Tables 1 and 2 for questions (b) and (d) suggests that most students who engaged with the QuILT answered question (d) correctly but struggled to calculate every single matrix element of the 6x6 matrix on the posttest in question (b).

Moreover, based on think-aloud interviews, we find that QM experts are more likely to consider whether different operators are diagonal in a given representation *before* choosing a basis to evaluate the matrix elements of the two Hamiltonians. They generally preferred to use the coupled representation since both Hamiltonians are diagonal in that representation (all off-diagonal matrix elements in questions (b) and (c) are zero). Since think aloud interviews suggest that students did not, in general, automatically do this type of metacognition before selecting a basis for evaluating the matrix elements, the CQS will be revised to explicitly offer such opportunity to students. In particular, more scaffolding will be provided to help students construct a set of basis states that is not only complete, but is also convenient for evaluating the matrix elements of operators corresponding to observables in question (e.g., choosing the coupled representation would have made both the Hamiltonians diagonal in the basis and made it significantly easier to calculate the matrix elements). We will refine the second section of the CQS, which deals with the coupled representation, to offer additional practice in constructing a basis in this less familiar case. Also, the third section of the CQS will be refined to offer more practice in identifying a representation in which a given operator is diagonal.

## IV. SUMMARY

The use of clicker questions is versatile and relatively easy to implement [1-2]. We describe the development, validation and in-class implementation of a CQS on addition of angular momentum that is inspired by the learning goals and guided inquiry-based sequences in a research-validated QuILT [5]. This CQS is composed of three sections: the first two focus on a review of the uncoupled and coupled representations for a system of two spin-1/2 particles, and the third is an extension to product spaces with higher dimension. The different sections of the CQS can be spread across separate lecture periods if needed, or can be implemented together, e.g., to review the concepts.

Development of a research-validated learning tool is an iterative process. After the in-class implementation of the CQS on the addition of angular momentum, we found that the CQS was effective in helping students construct a complete set of basis states in a product space and in calculating matrix elements for an operator that is diagonal in that basis. However, in-class evaluation also shows that further iterations are needed to guide students in selecting a representation that simplifies the task of calculating the matrix elements of an operator corresponding to an observable (e.g., choosing a basis in which the Hamiltonian operator is diagonal). Appropriate modifications are being made to the CQS so that these issues can be addressed in the future iterations and implementations.

## ACKNOWLEDGEMENT
We thank the NSF for awards PHY-1505460 and 1806691.